\documentclass{kluwer}    % Specifies the document style.
\input epsf.sty
\newdisplay{guess}{Conjecture}

\def\be{\begin{equation}}
\def\ee{\end{equation}}
\def\lsim{\lower 2pt \hbox{$\, \buildrel {\scriptstyle <}\over
         {\scriptstyle \sim}\,$}}

\begin{document}                                                                                   
\begin{article}
\begin{opening}         
\title{Pulsar Slot Gaps and Unidentified EGRET Sources} 
\author{Alice \surname{Harding}}  
\author{Alexander \surname{Muslimov}}
\runningauthor{Harding \& Muslimov}
\runningtitle{Pulsar Slot Gap Emission}
\institute{NASA Goddard Space Flight Center}
\date{June 30, 2004}

\begin{abstract}
A new picture of pulsar high-energy emission is proposed that is different from both the traditional polar cap and outer gap models, but combines elements of each.  The slot gap model is based on electron acceleration along the edge of the open field region from the neutron star surface to near the light cylinder and thus could form a physical basis for the two-pole caustic model of \cite{DR03}.  Along the last open field line, the pair formation front rises to very high altitude forming a slot gap, where the accelerating electric field is unscreened by pairs. The resulting radiation features both hollow cones from the lower-altitude pair cascades, seen at small viewing angles, as well as caustic emission on the trailing-edge field lines at high altitude, seen from both poles at large viewing angle.  
The combination of the small solid angle of slot gap emission ($\ll$ 1 sr) with a high probability of viewing the emission predicts that more gamma-ray pulsars could be detected at larger distances.  In this picture, many of the positional coincidences of radio pulsars with unidentified EGRET sources become plausible as real associations, as the flux predicted by the slot gap model for many of the pulsars would provide the observed EGRET source flux.  The expected probability of seeing radio-quiet gamma-ray pulsars in this model will also be discussed.
\end{abstract}
\keywords{neutron stars, pulsars, acceleration, gamma-rays}

\end{opening}           

\section{Introduction}  
                    % Produces section heading.  Lower-level
                    % sections are begun with similar 
                    % \subsection and \subsubsection commands.
The number of rotation-powered pulsars with detected emission at X-ray and $\gamma$-ray energies has been steadily growing \cite{Thomp01, BA02}.  Although 
the theory of pulsar acceleration and high-energy emission has been studied for over thirty years, the origin of the pulsed non-thermal emission is a question that remains unsettled.
The observations to date have not been able to clearly distinguish between an emission site
at the magnetic poles \cite{DH96} and emission in the outer magnetosphere \cite{CHR86, HS01}.   In the case of polar cap (PC) models, while the energetics and pair-cascade spectrum have had success in reproducing the observations, the predicted beam size of 
radiation emitted near the neutron star (NS) surface is too 
small to produce the wide pulse profiles that are observed \cite{Thomp01}.  However, \citeauthor{Arons83} (\citeyear{Arons83}) first noted the possibility of a high-altitude acceleration region or ``slot gap" along the last open field line where the electric field is lower.  The slot gap 
forms because the pair formation front, above which the accelerating field is screened, occurs at increasingly higher altitude as the magnetic colatitude approaches the edge of the open field region (\citeauthor{AS79}, \citeyear{AS79} 
[AS79]).  We have re-examined the slot-gap model (\citeauthor{MH03}, \citeyear{MH03} [MH03]) with the inclusion of two new features: 1) the acceleration due to inertial-frame dragging \cite{MT92} and 2) the additional decrease in the electric field near the boundary at the edge of the polar cap due to the narrowness of the slot gap.  These two effects combine to enable acceleration to altitudes 
approaching the light cylinder in the slot gap at all azimuthal angles around the polar cap.  These features result in the production of a larger high-energy emission beam with small solid angle, both favorable for producing high fluxes for $\gamma$-ray pulsars.

\section{Formation of the Slot Gap}

In the space-charge limited flow acceleration model AS79, electrons are freely emitted from the neutron star surface near the magnetic poles and accelerated in steady-state.  When the electrons achieve a sufficient Lorentz factor they radiate curvature and inverse Compton photons that can produce electron-positron pairs in the strong magnetic field.  Some of the positrons turn around and accelerate downward toward the neutron star, while the electrons accelerate upward.  This pair polarization screens the electric field above a pair formation front (PFF). These models assume a boundary condition that the accelerating electric field and potential are zero at the last open field line.  Near the boundary, the electric field is lower and a larger distance is required for the electrons to accelerate to the Lorentz factor needed to radiate photons energetic enough to produce pairs.  The PFF thus occurs at higher and higher altitudes as the boundary is approached and curves upward, approaching infinity and becoming asymptotically parallel  to the  last open field line.  If the electric field is effectively screened above the PFF, then a narrow slot surrounded by two conducting walls is formed (see Figure 1).   Within this slot gap, the electric field is further screened by the presence of the second conducting boundary and acceleration occurs at a reduced rate.  Pair production and pair cascades therefore do not take place near the neutron star surface in the slot gap, as do the pair cascades along field lines closer to the magnetic pole (core), but occur at much higher altitudes.

\begin{figure}  % Figure 1
\epsfysize=10cm % fix the y-dimension and scales x-dim. to y-dim.

\hspace{2.0cm} \vspace{0.0cm}
\epsfbox{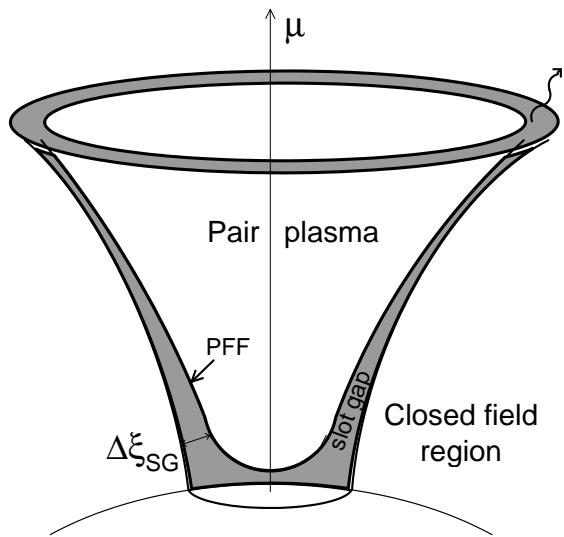}
%\vspace{10cm}  

\caption[h]{
Schematic illustration of polar cap geometry, showing the outer boundary of the open field line 
region (where $E_\parallel = 0$) and the curved shape of the pair formation front (PFF) which 
asymptotically approaches the boundary at high altitude.  The slot gap  exists between the pair plasma which results from the pair cascades above the PFF and the outer boundary.  A narrow
beam of high-energy emission originates from the low-altitude cascade on field lines interior 
to the slot gap.  A broader, hollow-cone beam originates from the high-altitude cascade above
the interior edge of the slot gap. $\Delta \xi_{_{\rm SG}}$ is the slot gap thickness (see text) and 
$\theta_{0, SG}$ is the colatitude at the center of the slot gap.
}
\end{figure}  

There are several important differences between the revised slot-gap model of MH03 and the original slot-gap model of AS79.  The inclusion of general relativistic frame dragging enables particle acceleration on both ``favorably" curved (curving toward the rotation axis) and ``unfavorable" curved (curving away from the rotation axis) field lines and also at all inclination angles.  MH03 also consider the radiation from pair cascades occurring along the interior edge of the slot gap.  The cascade radiation emission beam from the slot gap is thus a full hollow cone centered on the magnetic axis.  A narrower emission beam from field lines interior to the slot gap will form a core component of pairs and high-energy emission.  

Since the potential in the slot gap is uncreened, electrons on field lines which thread the slot 
gap will continue accelerating to very high altitudes.  \citeauthor{MH04}(\citeyear{MH04}, [MH04]) have derived 
the potential and accelerating electric field at high altitude in the extended slot gap by considering the effect of deviation of stream lines from the magnetic field lines of a static dipole.  In the absence of this effect, 
the difference between the actual charge density along field lines and the corotation (or Goldreich-Julian) charge density, which grows with altitude above the polar caps, would become
comparable to the Goldreich-Julian charge density itself, a situation which cannot be supported since it would disrupt the steady-state charge flow within the magnetic flux tube.  However, long before this would occur, the drift of 
electrons in the slot gap across the magnetic field will largely damp the growth of the charge
deficit and the large electric field which would be induced perdendicular to the magnetic field.
The residual parallel electric field is small and constant, but still large enough at all
altitudes up to nearly the light cylinder to maintain a flux of electrons with Lorentz factors exceeding
$10^7$ in the slot gap.

MH04 matched the high-altitude slot gap solution for the parallel electric field to the 
solution found at lower altitudes (MH03).  The result, for most inclination angles, 
is a continuously accelerating field from the neutron star surface to near the light cylinder 
along the last open field lines.  
The derived properties of the extended slot gap closely match the
geometrical properties required for the two-pole caustic model of Dyks \& Rudak (\citeyear{DR03}).
Such a model can reproduce the double-peaked profiles seen in many $\gamma$-ray pulsars like
the Crab, Vela and Geminga.  MH04 demonstrated through numerical simulation of slot gap
acceleration and radiation, that the extended slot gap radiation will produce the caustic peaks and 
pulse profiles similar to those of the two-pole caustic model.

\begin{figure}   % Figure 2

\epsfysize=15cm % fix the y-dimension and scales x-dim. to y-dim.
\hspace{-0.5cm}
\epsfbox{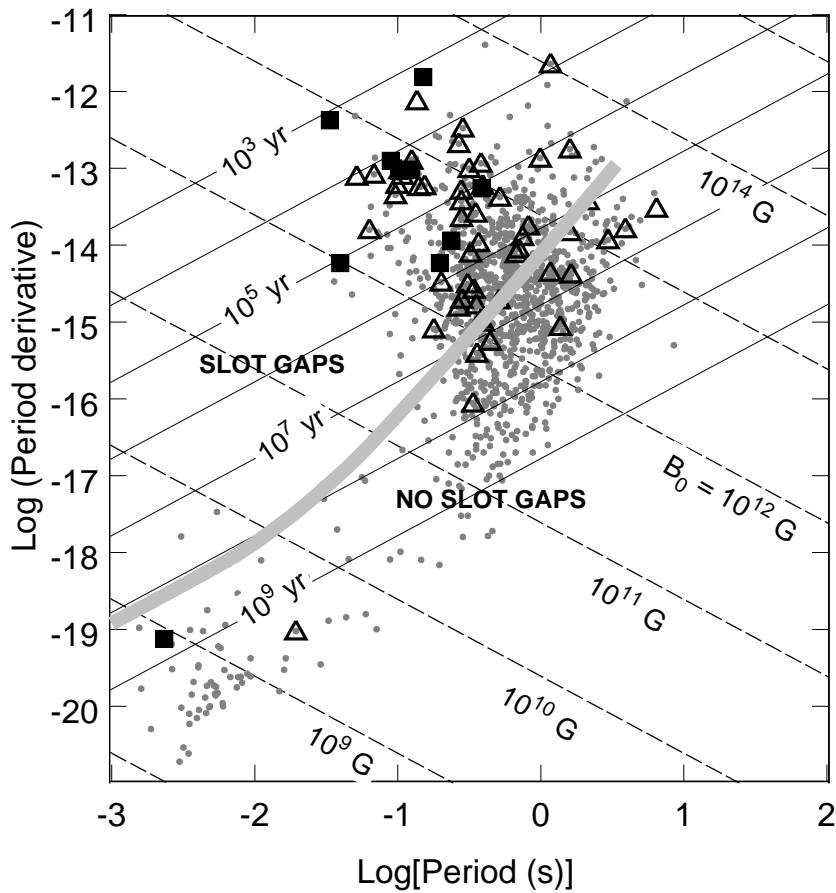}
\vskip -1.5cm
\caption{Radio pulsars from the ATNF catalog plotted on the $P$-$\dot P$ diagram.
The gray dots are radio pulsars, solid squares are EGRET pulsars, open triangles are
radio pulsars in EGRET sources error circles.  The gray line is the curvature radiation
pair death line.}
\end{figure}

The existence of a slot gap requires a dense enough pair plasma on interior field lines
to fully screen the parallel electric field, so that the inside walls of the gap have
vanishing electric field.  \citeauthor{HM02}(\citeyear{HM02}, HM02) found that pairs 
produced by CR are
always able to fully screen the $E_{\parallel}$ whereas pairs from ICS were only able to 
screen locally, if at all.  Therefore we can conclude that only those pulsars which can
produce CR pairs will have slot gaps.  Figure 2 reproduces the death lines for CR
pairs in $P$-$\dot P$ space from HM02, showing the region of pulsar parameter space required for
slot gap formation.  Generally, the younger pulsars, with ages less than $10^7$ yr and 
with higher magnetic fields fall into
this region, although one or two millisecond pulsars may also have slot gaps.

\section{Slot Gap Energetics}

\begin{figure}   % Figure 3

\epsfysize=12cm % fix the y-dimension and scales x-dim. to y-dim.

\hspace{-0.8cm} \vspace{0.0cm}
\epsfbox{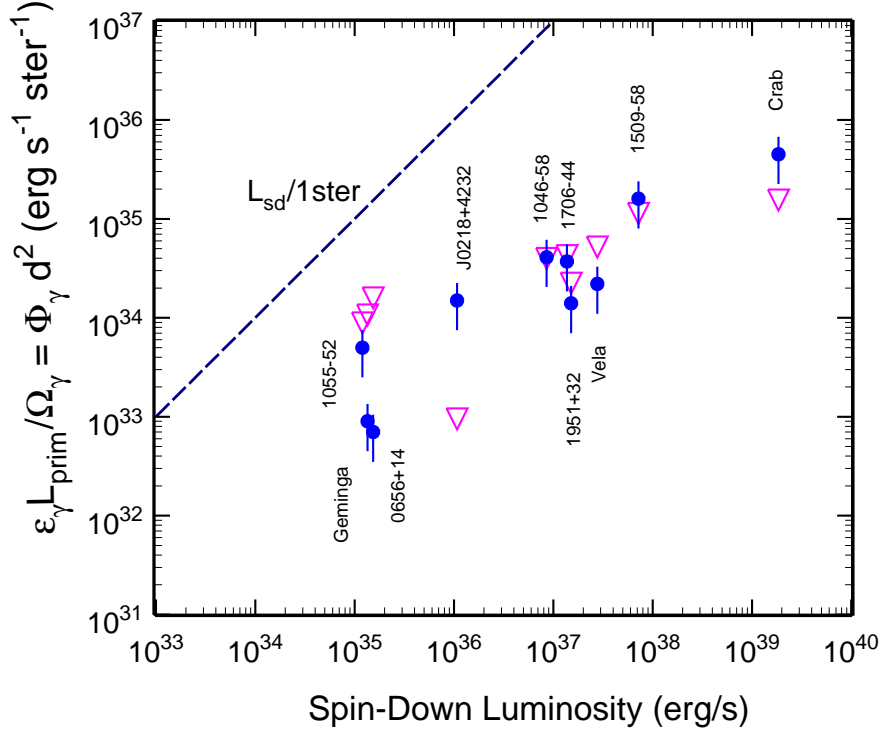}
%\vspace{9cm}

\caption[h]{
Observed flux above 1 keV, $\Phi_{\gamma}$, times distance squared (from \citeauthor{Thomp01}
(\citeyear{Thomp01}))
(solid circles) and theoretical values of specific high-energy luminosity from the slot gap, $\varepsilon _{\gamma }~L_{\rm prim}/ \Omega _{\gamma}$ from eq. (\ref{L/Omega1}) 
(upside-down triangles) vs. spin-down luminosity for known $\gamma$-ray pulsars.  An
efficiency of $\varepsilon_{\gamma} = 0.3$ was assumed. Also $\lambda = 0.1$, $\eta_{\gamma} = 3$ 
and the stellar parameters $R_6=1.6$ and $I_{45}=4$ were used. 
}
\end{figure}

The electrodynamics of the slot gap is primarily dependent on a single parameter, the slot gap width, $\Delta \xi_{_{\rm SG}}$.   The ratio of the electric field in the slot gap to the electric field in the core region of the PC is $E_{_{\rm SG}} / E_{\rm core} \propto \Delta \xi_{_{\rm SG}}^2/4$ and the luminosity of particles accelerated in the slot gap is $L_{\rm prim} \propto \Delta \xi_{_{\rm SG}}^3 L_{\rm SD}$ where  $L_{\rm SD}$ is the spin-down luminosity of the pulsar (for full details see \citeauthor{MH03}, \citeyear{MH03}).  
One can estimate the width of the slot gap as the magnetic colatitude where the variation in height of the 
curvature radiation PFF becomes comparable to a fraction $\lambda$ of the stellar radius $R$, or
\be  \label{SGwidth}
\Delta \xi _{_{\sc SG}} \approx
\left\{
\begin{array}{lr} 
 0.2~P_{0.1}(\lambda B_{12})^{-4/7}I_{45}^{-3/7}.
\label{deltaxi1} &  B_{12} \lsim 4.4 \\
0.1~P_{0.1}(\lambda B_{12}^{3/4})^{-4/7}
I_{45}^{-3/7} & B_{12} \gsim 4.4
\end{array} 
\right.
\ee
$I_{45}=I/10^{45}$ g$\cdot $cm$^2$, 
$I$ is the neutron star moment of inertia, $P_{0.1} = P/0.1$ s and $B_{12} = B_0/10^{12}$ G are the neutron star 
rotation period and surface magnetic field.  Here, $\Delta \xi _{_{\sc SG}}$ is in units of the polar cap
half-angle, $\theta_0 = \sin^{-1}(2\pi R/Pc)^{1/2}$.
The emission solid angle of radiation from the slot gap can be estimated by integrating over the thin annulus defined by the slot gap width (Eqn [\ref{SGwidth}]).  
\be \label{Omega}
\Omega _{SG} \approx {9\over 2}\pi \theta _0^2 \eta \Delta \xi _{_{\sc SG}}~~~~{\rm ster},
\ee
where $\eta \equiv r/R$ is the dimensionless radius of emission.  
Electrons accelerating in the slot gap will radiate curvature-radiation $\gamma$-rays,
becoming radiation-reaction limited at Lorentz factors
\be
\gamma \approx 3 \cdot 10^7 \left[
\kappa _{0.15} B_{12} {{R_6^3} \over P_{0.1}} {\eta \over \eta _{lc}} 
\nu _{_{\rm SG}} |\cos \chi |\right]^{1/4} ,
\label{gamma-CRRL}
\ee
where $\kappa _{0.15} = \kappa /0.15 \approx I_{45}/R_6^3$, $\nu _{_{\rm SG}} = 0.25 \Delta \xi _{_{\sc SG}}^2$, 
$R_6 = R/10^6$ cm is the neutron star radius, $\chi$ is the pulsar inclination angle 
and $\eta_{lc}$ is the dimensionless light cylinder radius.
Based on the luminosity of the primary electrons and the above solid angle estimate, we may derive the quantity, 
\begin{eqnarray} \label{L/Omega1}
L_{SG}(\Omega _{\gamma }) = {{\varepsilon _{\gamma }~
L_{\rm prim}}\over {\Omega _{SG}}}~~~~~~~~~~~~~~~~~~~~~~~~~~~~~~~~~~~~~~~~~~~~~~~~~~~~~~~~~~~~~~~ \nonumber \\
= 3\times 10^{34} ~\varepsilon _{\gamma }~ 
L_{\rm sd, 35}^{3/7} P_{0.1}^{5/7} R_6^{17/7} 
\eta^{-1}\lambda^{-8/7}\I_{45}^{-6/7}\kappa_{0.15}\cos^2\chi
~~{\rm erg\, s^{-1}\, ster ^{-1}}
\end{eqnarray}
where $L_{\rm prim}$ is the luminosity in primary electrons accelerated in the slot gap, 
$\varepsilon_{\gamma}$ is the 
radiation efficiency and $L_{\rm sd,35} \equiv L_{\rm sd}/10^{35}\,\rm erg s^{-1}$ is the spin-down luminosity.
The above expression for $L_{\gamma }(\Omega _{\gamma })$ is equivalent to the observed quantity 
$\Phi _{\gamma }~d^2$, where $\Phi _{\gamma }$ is the high-energy bolometric flux observed at the Earth, and $d$ is the distance to the pulsar. 
Figure 3 shows the observed (solid circles with error bars) and theoretical (upside-down triangles) 
values of $\Phi _{\gamma }d^2$ as a function 
of spin-down luminosity, $L_{\rm sd}$ for the 10 known $\gamma$-ray pulsars. 
The theoretical values are calculated for the parameters $\varepsilon _{\gamma } = 0.3$ and $\lambda = 0.1$ (see eq. 
[\ref{L/Omega1}]). Note that parameter $\varepsilon _{\gamma }$ can 
range from 0.2 to 0.5 in cascade simulations, and $\eta = 3$. 
In Figure 3 the dashed line represents the 
limit, where the spin-down luminosity is radiated into the unit 
solid angle, i.e. where 
$\Phi _{\gamma }d^2 = L_{\rm sd} / 1~{\rm ster}$. One can see that there is good agreement for most
high-energy pulsars except several of the pulsars, Geminga and PSR B0656+14, having low $L_{\rm sd}$, and for J0218+4232, which is a millisecond pulsar.   These pulsars are near or below the curvature radiation pair death line (see Figure 2 and HM02), and therefore have either very wide slot gaps or no slot gaps at all. 
All other high-energy pulsars depicted in Figure 3 
are well above the curvature radiation death lines and are expected to have slot gaps.

\section{Slot Gap Emission and EGRET Unidentified Sources}

\begin{figure}   % Figure 4

\epsfysize=10cm % fix the y-dimension and scales x-dim. to y-dim.

\hspace{-1.0cm}\vspace{0.0cm}
\epsfbox{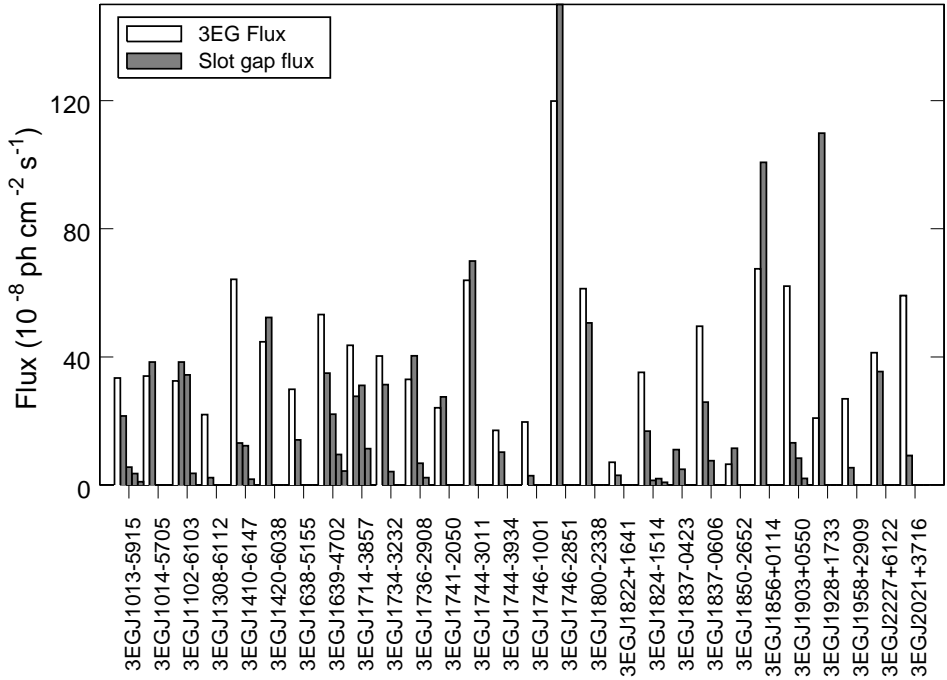}
\caption{Comparison of fluxes of EGRET sources (white bars) and of associated radio pulsars 
(gray bars) as predicted by the slot gap
model (Eqn [\ref{L/Omega1}] and \protect\citeauthor{MH03} \protect\citeyear{MH03})}
\end{figure}

There are presently 53 radio pulsars in the error circles of 28 EGRET unidentified sources
\cite{Kramer03, Grenier04}.  Nearly all of these pulsars have been 
discovered by surveys such as the Parkes Multibeam \cite{Man01} and by deep
pointed searched by Arecibo and GBT radio telescopes \cite{Camilo04} after the end of EGRET operation.
The pulsars coincident with EGRET sources are plotted in $P$-$\dot P$ space in Figure 2.
Most of them are young ($\tau \lsim 10^7$ yr), high-field pulsars 
above the CR pair death line and many are expected to have slot gaps. 
There are unfortunately too few photons in any of the sources to do credible pulsation 
searches of the EGRET 
archival data, so such searches must await AGILE or GLAST.  Many of the coincident EGRET 
sources contain multiple radio pulsars within the error circles.  
These are shown in Figure 4 with their
observed average fluxes from the 3rd EGRET catalog \cite{Hart99}.

We have computed the $\gamma$-ray fluxes predicted by the slot gap model (Eqn [\ref{L/Omega1}]) 
for the EGRET 
source-coincident radio pulsars and the results are plotted in Figure 4.  The fluxes of 
the pulsars, $L_{SG}(\Omega _{\gamma})/d^2$, where $d$ is the distance, are shown as dark bars alongside 
the coincident 
EGRET source fluxes shown as light bars.  In about 18-22 out of the 28 sources, the predicted
slot gap flux from either an individual pulsar or several pulsars combined could account
for the EGRET source flux.  Thus more than two thirds of these associations are 
physically plausible,
making the pulsars viable counterparts for the EGRET sources.  In contrast, the predicted
fluxes from the standard polar cap model (e.g. HM02), assuming a solid angle of 1 sr., 
would be comparable to the EGRET source fluxes in only about 5 of the cases.

\section{Conclusions}

In the slot gap model, pulsar high-energy emission comes from high altitudes, between
2-3 stellar radii above the neutron star surface to nearly the light cylinder.
Two resulting characteristics of radiation from the slot gap: small solid angle and a
wide emission beam,
combine to provide a significantly larger flux for $\gamma$-ray pulsars than in the
standard polar cap model.  The small solid angle allows the radiation from $\gamma$-ray pulsars to be seen 
by any given detector at larger distances, and many of the spatially coincident pulsars are at large
distances, as determined from their dispersion measure.  
Thus, a larger number of the recently detected radio pulsars
that are in or near EGRET source error circles become plausible candidates for the $\gamma$-ray
sources.  It is possible that half of
the non-variable unidentified EGRET sources in the galactic plane are radio loud $\gamma$-ray
pulsars.  A population synthesis study of radio-loud and radio-quiet pulsars in the
galaxy \cite{Gonthier04a, Gonthier04b} have found that in the slot gap model one expects
that many more radio-loud than radio-quiet pulsars are counterparts to EGRET sources.
This result is due to the fact that the radio emission occurs along the same open field lines
as the $\gamma$-ray emission, in contrast to outer gap models where the radio emission is
assumed to originate from open field lines on the opposite pole of the neutron star from the
high-energy emission.  The ratio of radio-loud to radio-quiet $\gamma$-ray pulsars detected by EGRET,
and eventually by GLAST, will be a good discrimator between emission models.

\end{article}
\end{document}